%
%
\hsize=15.truecm\vsize=23.truecm\parindent=2.5mm\parskip 1pt plus 1pt
\font\titolo=cmbx10 scaled \magstep1
\let\0=\noindent\def\*{\vskip0.3truecm plus0.1truecm}
\def\fra#1#2{{#1\over#2}}\def\media#1{\langle{#1}\rangle}
\let\dpr=\partial\let\ig=\int\def\\{\hfill\break}
\def\annota#1#2{\footnote{${}^#1$}{#2}}
\def\V#1{{\,\underline#1\,}}\let\==\equiv\def\ie{{\it i.e.\ }}
\let\ciao=\bye

\newdimen\xshift \newdimen\xwidth \newdimen\yshift
\def\ins#1#2#3{\vbox to0pt{\kern-#2 \hbox{\kern#1
#3}\vss}\nointerlineskip} \def\eqfig#1#2#3#4#5{ \par\xwidth=#1
\xshift=\hsize \advance\xshift by-\xwidth \divide\xshift by 2
\yshift=#2 \divide\yshift by 2 \line{\hglue\xshift \vbox to #2{\vfil #3
\includegraphics{#4.ps} }\hfill\raise\yshift\hbox{#5}}}

\def\8{\write13}



\frenchspacing
\centerline{\titolo M\'ecanique statistique hors \'equilibre~: l'h\'eritage
de Boltzmann}

\*
\centerline{\it Giovanni Gallavotti}
\centerline{Dipartimento di Fisica, Universit\'a di Roma La Sapienza}
\*
\*
{\it Resum\'e~: On place quelques applications r\'ecentes de la th\'eorie du
Chaos en perspective avec la th\'eorie classique de Boltzmann.}
\*

\def\PP{{\cal P}}\let\Eqa=\eqno\def\equ{}\def\;(#1){#1}
Boltzmann entreprit, \;([B66]), de prouver l'existence des atomes en
poursuivant un programme d\'ej\`a amorc\'e par ses
pr\'ed\'ecesseurs. Son approche \'etait d'\'etablir que la conception
de la mati\`ere en tant qu'agglom\'eration d'atomes ob\'eissants aux lois
de la m\'ecanique conduisait \`a la d\'eduction des propri\'et\'es de
la mati\`ere que connaissaient alors les exp\'erimentateurs et les
th\'eoriciens.

Ainsi Boltzmann produisit des versions de plus en plus raffin\'ees du
{\it th\'eor\`eme de la chaleur}, \;([B68]), \;([B71]), \;([B72]),
\;([B77]). Au
d\'ebut il s'agissait de faire voir qu'il est possible de d\'efinir des
quantit\'es m\'ecaniques associ\'ees, par exemple, \`a un gaz enferm\'e dans
un conteneur cubique de volume $V$, telles que~:
\smallskip
\halign{\kern1cm #\ =&{\it\ #}\hfill\cr
$T$& \'energie cin\'etique moyenne\cr
$U$& \'energie totale\cr
$V$& volume\cr
$p$& impulsion moyenne transf\'er\'ee aux parois par collision et
par unit\'e de surface\cr}
\smallskip
\0o\`u les moyennes sont calcul\'ees empiriquement en supposant les
particules ind\'ependantes et \`a distribution uniforme sur une sph\`ere dan=
s
l'espace des impulsions et dans le volume $V$ des positions.

Le th\'eor\`eme \`a prouver est alors que si on varie $U$ et $V$ de
$dU$ et $dV$ et que l'on calcule la quantit\'e~:

$$\fra{dU\,+\,p\,dV}{T}$$

\0o\`u $p$ et $T$ d\'ependent de $U$ et $V$, on trouve
{\it une diff\'erentielle exacte}~:
c'est \`a dire qu'il existe une fonction $S(U,V)$ telle que
$dS=\fra{dU\,+\,p\,dV}{T}$.
\medskip

Suite aux travaux de Boltzmann, Helmholtz consid\'era les syst\`emes
m\'ecaniques {\it monocycliques}, c'est-\`a-dire les syst\`emes dont
tout mouvement d'\'energie donn\'ee est p\'eriodique et non d\'eg\'en\'er\'e
(ce qui veut dire que les mouvements d'\'energie donn\'ee ne
diff\`erent entre eux que par un d\'ecalage du temps d'observation),
\;([H]).

Il fit voir que, en g\'en\'eral, si on imagine que les mouvements (``{\it
\'etats}'') de tels syst\`emes sont param\'etr\'es par leur
\'energie totale $U$ et par un param\`etre $V$ dont les potentiels
$\varphi_V$ des forces qui agissent sur le syst\`eme d\'ependent, alors en
d\'efinissant~:
\smallskip
\halign{\kern1cm #\ =&{\it\ #}\hfill\cr
$T$& \'energie cin\'etique moyenne\cr
$U$& \'energie totale\cr
$V$& volume\cr
$p$& $\media{-\fra{\dpr}{\dpr V}\varphi} $\cr}
\smallskip
\0o\`u $\media{F}$ maintenant d\'enote pr\'ecis\'ement la moyenne de
$F$ par rapport au temps (et donc n'est pas d\'efinie empiriquement
comme dans le cas pr\'ecedent), on trouve {\it en g\'en\'eral}~:

$$\fra{dU\,+\,p\,dV}{T}=\ \hbox{\rm diff\'erentielle exacte}\Eqa(1)$$

\0Celle-ci aurait pu n'\^etre rien de plus qu'une curiosit\'e. Mais
Boltzmann avait une conception {\it discr\`ete} de la nature~: m\^eme
s'il ne l'avait pas explicitement dit dans ses \'ecrits populaires, on
le verrait dans ses travaux scientifiques o\`u l'emploi de l'analyse,
avec ses int\'egrales et ses d\'eriv\'ees, est souvent vu comme un moyen
technique pour venir \`a bout du calcul de sommes et de diff\'erences,
\;([B05]).

Donc pour Boltzmann le mouvement n'est qu'une \'evolution discr\`ete
o\`u l'espace des phases est quadrill\'e en petites cellules \`a $6N$
dimensions ($N$ \'etant le nombre de mol\'ecules) dont {\it une} contient
le point qui repr\'esente l'\'etat instantan\'e du syst\`eme.
L'\'evolution apparait comme les d\'eplacements successifs du point
repr\'esentatif d'une cellule \`a une autre, alors que le temps
s'\'ecoule d'une petite quantit\'e discr\`ete $h$. Bien s\^ur le
d\'eplacement doit \^etre conforme aux lois du mouvement.

C'est une repr\'esentation tr\`es famili\`ere aujourd'hui \`a qui
essaye de simuler sur ordinateur les mouvements d'un gaz de
particules. Sur l'ordinateur les \'etats microscopiques du gaz sont
repr\'esent\'es par des cellules (car les coordonn\'ees des points
sont repr\'esent\'ees par des nombres qui sont d\'e\-ter\-mi\-n\'es avec une
pr\'ecision qui est loin d'\^etre infinie et qui d\'epend de la
machine ou plut\^ot du logiciel que l'on emploit) et l'\'evolution se
d\'eroule par pas discrets; le programme qui effectue ces pas est
\'ecrit avec les lois du mouvement comme guide.

De ce point de vue le mouvement est une permutation des cellules qui
repr\'esentent l'\'etat microscopique. Le syst\`eme est alors toujours
en \'evolution p\'eriodique~: car toute permutation d'un nombre fini
d'objets (les cellules d'\'energie totale $U$ donn\'ee, dans le cas
pr\'esent) engendre une \'evolution cyclique.

On imagine que l'on fixe l'\'energie totale $U$ et que les forces
agissantes sur le syst\`eme sont param\'etr\'ees par le volume $V$~:
en fait on imagine que, quoi que l'on fasse, les forces entre les
particules ne varient pas et seules les forces entre les particules et
les parois peuvent changer (\`a cause des mouvements des parois et des
changements de volume qui en d\'ecoulent).

Alors l'hypoth\`ese de monocyclicit\'e de Helmholtz, de
non-d\'eg\'en\'erescence des mouvements d'\'ener\-gie donn\'ee,
correspondrait
\`a dire que l'\'evolution est une permutation \`a un seul cycle des
cellules et donc on serait dans la situation o\`u le syst\`eme est
monocyclique~: cette hypoth\`ese est connue comme l'{\it hypoth\`ese
ergodique}.

Sous cette hypoth\`ese on devrait avoir la possiblit\'e de trouver, {\it
en g\'en\'eral}, une analogie m\'ecanique de la thermodynamique et un
th\'eor\`eme g\'en\'eral de la chaleur. Puisque les moyennes doivent se
calculer, alors, par la distribution uniforme sur l'espace des
cellules d'\'energie donn\'ee (car les cellules ont des tailles \'egales) on
se trouve oblig\'e de v\'erifier que dans le cas d'un gaz (ou m\^eme d'un
liquide ou d'un solide, vue la g\'en\'eralit\'es des consid\'erations en
question)
$$\fra{dU\,+\,p\,dV}{T}\quad\hbox{est exact si}\quad
p=-\media{\fra{\dpr}{\dpr V}
\varphi_V}.$$
\indent
Cette propri\'et\'e, cas particulier d'une propri\'et\'e plus g\'en\'erale
que Boltzmann appella {\it orthodicit\'e}, doit \^etre accompagn\'ee par
la propri\'et\'e suppl\'ementaire que $p$ est {\it aussi} l'impulsion moyenn=
e
transf\'er\'ee aux parois par les collisions, par unit\'e de temps et de
surface. Si cela est bien le cas on aura prouv\'e que en g\'en\'eral un
th\'eor\`eme de la chaleur est valable.

C'est ce que Boltzmann fit en 1884, \;([B84]), en fondant, en m\^eme
temps, la {\it th\'eorie des ensembles statistiques} (qui est souvent
attribu\'ee \`a Gibbs, mais pas par Gibbs lui m\^eme, \;([G])).

Il est tout \`a fait remarquable que le th\'eor\`eme de la chaleur,
eq.\equ(1), est valable tant pour les petits syst\`emes (m\^eme \`a une
particule, si la non-lin\'earit\'e du mouvement est suffisante de fa{\c c}on
\`a rendre l'hypoth\`ese ergodique raisonnable) que pour les grands
(avec $10^{23}$ particules).

\medskip\centerline{{\it L'exactitude de
$\displaystyle\fra{dU\,+\,p\,dV}{T}$ ne d\'epend pas de la taille du
syst\`eme},
\;([B84]).}
\medskip

Cette ind\'ependance est d'ailleurs une propri\'et\'e absolument
fondamentale et elle permit \`a Boltzmann de se d\'egager des
critiques qui lui \'etaient adress\'ees.

Les critiques, par Zermelo et m\^eme par Poincar\'e, \'etaient
subtiles et portaient sur le principe selon lequel il serait
impossible de d\'eduire les lois macroscopiques (irr\'eversibles) d'une
m\'ecanique r\'eversible qui est n\'ecessairement cyclique et donc
apparemment pas irr\'eversible (au bout d'un temps de {\it r\'ecurrence}
le syst\`eme revient \`a son \'etat initial, contre toute intuition sur
le comportement des syst\`emes macroscopiques), \;([B96]), \;([B97]).

Ces critiques s'adressaient surtout \`a l'\'equation de Boltzmann et par
cons\'equent \`a l'approche irr\'eversible \`a l'\'equilibre. Boltzmann,
comme il est bien connu, r\'epondit qu'on ne pouvait pas ne pas tenir compte
des \'echelles de temps n\'ecessaires \`a r\'eveler des contradictions. Pour
voir, au niveau macroscopique, les effets de la r\'eversibilit\'e
microscopique, le temps qu'il fallait attendre \'etait \'enorme qu'on le
mesure en heures ou en \^ages de l'Univers,
\;([B02]).  Apr\`es quoi on observerait une \'evolution anormale pour
revenir presque imm\'ediatement au comportement normal et pour une
p\'eriode de dur\'ee encore aussi longue.

Mais cet argument, \`a la d\'efense de l'\'equation de Boltzmann,
d\'etruisait aussi apparemment la signification du th\'eor\`eme de la
chaleur et la possibilit\'e de d\'eduire la thermodynamique de la
m\'ecanique et de l'hypoth\`ese ergodique. Car pour que le
th\'eor\`eme de la chaleur ait un int\'er\^et quelconque il faut que les
moyennes dont il parle soient atteintes dans un laps de temps
raisonnablement court~: mais si le temps de r\'ecurrence (c'est \`a
dire le temps n\'ecessaire au point repr\'esentatif du syst\`eme pour
revenir \`a la cellule initiale dans l'espace des phases) est \'enorme
alors les moyennes des observables risquent d'\^etre atteintes sur un
temps du m\^eme ordre, ce qui signifierait qu'elles n'ont pas d'int\'er\^et
physique.

Boltzmann aper{\c c}ut cette difficult\'e et fut conduit \`a dire que
dans un syst\`eme macroscopique tout se passe comme si les moyennes
sur des temps courts \'etaient les m\^emes que sur les temps
(inobservables) de r\'ecurrence. Ceci serait d\^u au fait que si le
nombre de particules est tr\`es grand, les grandeurs d'int\'er\^et
thermodynamique prennent la m\^eme valeur sur presque tout l'espace des
phases~: ce qui leur permet d'atteindre leur valeur moyenne sur des
temps tr\`es courts qui n'ont rien \`a voir avec le temps de
r\'ecurrence (qui est infini \`a tout point de vue). Elle prennent
la m\^eme valeur parce qu'elles sont \`a leur tour des moyennes sur les
particules et ne d\'ependent pas de l'\'etat de particules
individuelles.

Donc l'hypoth\`ese ergodique sugg\`ere l'ensemble microcanonique pour
le calcul des moyen\-nes~: c'est un fait g\'en\'eral que ces moyennes
v\'erifient les relations thermodynamiques qui, d'un autre c\^ot\'e, sont
observables gr\^ace \`a la lois des grands nombres qui fait que ces
grandeurs ont la m\^eme valeur partout (ou presque) dans l'espace
des phases.

Il s'en suit que l'hypoth\`ese ergodique n'est pas une justification
de la thermodynamique et ne joue qu'un r\^ole cin\'ematique. La
thermodynamique est une {\it identit\'e} m\'ecanique qui devient
observable au niveau macroscopique gr\^ace \`a la loi des grands nombres,
\;([Ga1]).

\*
Une fois achev\'ee cette admirable construction conceptuelle on se pose
la question de savoir si on peut faire de m\^eme dans le cas des syst\`emes
{\it hors \'equilibre}.

Ce sont des syst\`emes de particules sur lesquels agissent une ou
plusieurs forces conservatives dont le travail est dissip\'e dans des
thermostats, permettant ainsi au syst\`eme d'atteindre un \'etat
stationnaire.

C'est un probl\`eme pas vraiment touch\'e par Boltzmann qui \'etudia
en d\'etail le probl\`eme du retour \`a l'\'equilibre d'un gaz
perturb\'e de son \'etat d'\'equilibre (retour qui se d\'eroule selon
l'\'equation de Boltzmann). Et il peut para{\^\i}tre \'etrange qu'un
probl\`eme si naturel et d'une telle importance soit rest\'e
essentiellement ouvert jusqu'\`a nos jours.

On remarque imm\'ediatement une profonde diff\'erence par rapport au
probl\`eme de la th\'eorie des \'etats d'\'equilibre~: il n'y a pas une
v\'eritable th\'eorie macroscopique (comparable \`a la thermodynamique
classique) qui puisse servir de guide et qui fournisse des r\'esultats
\`a prouver.

Une diff\'erence technique importante est que l'on peut s'attendre \`a ce qu=
e
le comportement physique du syst\`eme d\'epende de la m\'ethode qu'on
emploie pour enlever la chaleur produite par le travail des forces qui
agissent. Ce qui peut donner le souci qu'une th\'eorie g\'en\'erale
soit impossible \`a cause de la grande vari\'et\'e de forces
thermostatiques qu'on peut imaginer pour un m\^eme syst\`eme.

Mais, \`a mon avis, il ne s'agit que d'une difficult\'e apparente qui
disparait au fur et \`a mesure qu'on pr\'ecise la th\'eorie.

Donc on va imaginer un syst\`eme de particules sur lesquelles agissent
des forces externes non conservatives et un m\'ecanisme quelconque
qui emp\`eche le r\'echauffement. On va mod\'eliser ce thermostat par
des forces additionnelles. Par exemple, si le syst\`eme est un gaz de
sph\`eres dures enferm\'ees dans un conteneur p\'eriodique avec
quelques obstacles fixes et soumises \`a un champ de force $\V E$, on
peut imaginer que les \'equations du mouvement soient~:

$$m \ddot{\V x}_i=\V f_i+\V E-\nu\dot{\V x}_i=\Phi_i(\V x,\dot{\V
x})\Eqa(2)$$
o\`u les $\V f_i$ sont les forces entre particules (sph\`eres dures
\'elastiques) et entre particules et obstacles (qui sont aussi des sph\`eres
dures \'elastiques).

Ici $\nu \dot{\V x}=\nu(\dot{\V x})\, \dot{\V x}$  est le mod\`ele de
thermostat. La vraie {\it difficult\'e} est que l'\'evolution engendre une
contraction du volume de l'espace des phases car~:

$$\fra{d}{dt}(d\V x \,d\dot{\V x})= {\rm div}\Phi\,\cdot\,(d\V x
\,d\dot{\V x})\Eqa(3)$$
et la dissipativit\'e entraine $-\media{{\rm div}\Phi}>0$, et donc l'\'etat
stationnaire devra \^etre une distribution de probabilit\'e $\mu(d\V
x,d\dot{\V x})$ {\it concentr\'ee sur un ensemble de volume nul}. Elle
ne pourra pas \^etre d\'ecrite par une densit\'e de la forme~: $\rho(\V
x,\dot{\V x})\,d\V x,d\dot{\V x}$.

Du coup on ne peut m\^eme pas \'ecrire les formules qui expriment
formellement les moyennes des observables par rapport \`a l'\'etat
stationnaire en termes d'une fonction de densit\'e inconnue.

N\'eanmoins on voudrait avoir de telles expressions pour pouvoir esp\'erer
en tirer des cons\'e\-quen\-ces g\'en\'erales, du type du th\'eor\`eme de la
chaleur, qui puissent \^etre observ\'ees dans les petits syst\`emes
(parce que directement observables) et dans les grands aussi (pour des
raison diff\'erentes).

L'id\'ee clef a pris forme au d\'ebut des ann\'ees 1970, 1973 au plus
tard, et est due \`a Ruelle~: mais dans un contexte apparemment assez
diff\'erent du n\^otre (celui de la m\'ecanique des fluides et de la
turbulence). On con{\c c}oit les mouvements turbulents d'un fluide
stationnaire ou d'un gaz de particules comme des {\it mouvements
chaotiques}.
\medskip

Cela ne demande pas \`a premi\`ere vue beaucoup d'imagination~: mais le poin=
t
est que l'hypoth\`ese est pos\'ee dans un sens technique pr\'ecis, \;([R1]),
\;([R2]). Dans l'interpr\'etation d'auteurs successifs, \;([GC]), on dit
que le principe est que {\it le syst\`eme est ``hyperbolique'' ou d'
``Anosov''}. C'est l'{\it hypoth\`ese chaotique}.
\medskip

Cela veut dire que en tout point $x$ de l'espace des phases on peut
\'etablir un syst\`eme {\it covariant} de coordonn\'ees locales tel
que l'\'evolution temporelle $n\to S^nx$ observ\'ee dans ce syst\`eme
voit $x$ comme un point fixe (car on le suit) hyperbolique.
C'est-\`a-dire on voit depuis $x$ les autres points bouger de la m\^eme
fa{\c c}on qu'on les voit si on regarde les mouvements \`a partir du point
fixe instable d'un pendule~: la diff\'erence \'etant que cela est vrai pour
tout point (et non pas pour un point isol\'e comme dans le cas du
pendule).

On aura cette propri\'et\'e valable \`a l'\'equilibre aussi bien que
hors \'equilibre~: les mouvements des mol\'ecules sont chaotiques m\^eme
dans les \'etats d'\'equilibre. Pour comprendre ce qui se passe il
convient de revenir au point de vue discret de Boltzmann.

Si un syst\`eme est dissipatif il y a des difficult\'es
suppl\'ementaires car il est clair qu'on a beau rendre petites les
cellules de l'espace des phases, on n'arrivera jamais \`a un
syst\`eme dynamique discret qui puisse \^etre d\'ecrit comme une
permutation des cellules~: la contraction de l'espace des phases
entraine que certaines cellules ne seront jamais plus visit\'ees
m\^eme si on les a visit\'ees au d\'epart (par exemple parce que l'on a
initi\'e le mouvement \`a partir d'elles). Les mouvements se
d\'eroulent asymptotiquement sur un {\it attracteur} (qui est plus
petit que tout l'espace des phases, bien que si on consid\`ere
l'espace de phases comme continu l'attracteur pourrait \^etre
dense\annota{1}{Ce qui montre seulement que la notion de ``grandeur''
d'un attracteur est plut\^ot d\'elicate}).

Mais si on consid\`ere seulement les cellules sur lesquelles se
d\'eroule le mouvement on est dans une situation {\it identique \`a
l'\'equilibre et hors \'equilibre}. On imagine que le mouvement est une
permutation \`a un cycle, et donc il y aura un \'etat stationnaire unique.
Le temps pour parcourir le cycle sera, bien \'evidemment, toujours du m\^eme
ordre de grandeur qu'\`a l'\'equilibre (dans des situations pas trop
extr\^emes des param\`etres qui d\'eterminent les forces agissantes sur le
syst\`eme)~: donc la raison pour laquelle on peut esp\'erer observer les
moyennes temporelles et les calculer par int\'egration par rapport \`a une
distribution de probabilit\'e sur l'espace des phases reste la m\^eme que
celle d\'ej\`a discut\'ee dans le cas d'\'equilibre (et li\'ee \`a la loi
des grands nombres).

Toutefois il y a une difficult\'e~: c'est une difficult\'e qu'on aurait pu
discuter d\'ej\`a dans le cas de l'\'equilibre. On a suppos\'e, sans
critique, que les cellules de l'espace des phases \'etaient toutes
\'egales. Mais m\^eme dans le cas de l'\'equilibre les syst\`emes sont
chaotiques et donc toute cellule est d\'eform\'ee par l'\'evolution
temporelle qui la dilate dans certaines directions et la contracte
dans d'autres.

Il appara{\^\i}t alors que la repr\'esentation du mouvement comme \'evolutio=
n
d'une cellule vers une autre de forme et de taille identique est loin
d'\^etre triviale. Elle est en fait une hypoth\`ese {\it forte} sur la
dynamique, qui, \`a l'\'equilibre, s\'electionne l'ensemble
microcanonique comme distribution correcte \`a utiliser pour calculer
les moyennes temporelles (et qui entraine le th\'eor\`eme de la
chaleur). Il y a bien d'autres distributions invariantes sur
l'espace des phases (contrairement \`a ce qu'on entend dire parfois) et
l'hypoth\`ese apparemment innocente que le mouvement se repr\'esente
comme une permutation de cellules identiques en s\'electionne une
particuli\`ere.

Hors \'equilibre la difficult\'e devient plus manifeste. Car le
volume des cellules ne reste m\^eme pas invariant contrairement au
cas de l'\'equilibre (gr\^ace au th\'eor\`eme de Liouville). De plus hors
\'equilibre il faut s'attendre \`a ce que la repr\'esentation du mouvement
comme
\'evolution de cellules identiques conduise \`a s\'electionner une
distribution de probabilit\'e particuli\`ere sur l'espace des phases,
concentr\'ee sur les cellules qui constituent l'attracteur, \;([Ga2]).

L'int\'er\^et et l'importance des syst\`emes chaotiques au sens de
l'hypoth\`ese chaotique est que, en effet, pour tous ces syst\`emes
{\it il y a une unique distribution stationnaire $\mu$ sur l'espace
des phases qui donne les moyennes des grandeurs observ\'ees sur les
mouvements qui commencent dans la grande majorit\'e des cellules
identiques en lesquelles on peut imaginer de diviser l'espace des
phases}. C'est un r\'esultat fondamental d\^u \`a Sinai et \`a
Ruelle--Bowen~: ainsi la distribution $\mu$ s'appelle {\it distribution
SRB}, \;([S]), \;([BR]). Dans le cas de l'\'equilibre, elle co{\"\i}ncide
avec la distribution microcanonique.

Ce n'est pas ici le lieu de poursuivre la critique de la vision
discr\`ete du mouvement, bien qu'elle soit int\'eressante ne fusse que
pour une interpr\'etation correcte des simulations num\'eriques qui se font
de plus en plus fr\'equentes, voir la note ${}^2$ page suivante.

L'hypoth\`ese chaotique conduit naturellement \`a une
repr\'esentation discr\`ete diff\'erente du mouvement qui non seulement ne
souffre pas des critiques qu'on vient de mentionner, mais qui nous donne une
formule explicite pour la valeur des moyennes des observables, valable \`a l=
a
fois \`a l'\'equilibre (o\`u elle se r\'eduit \`a l'ensemble microcanonique)
et hors \'equilibre.

Cette nouvelle repr\'esentation est aussi bas\'ee sur des cellules~: mais
elle ne sont pas vraiment petites dans le sens qu'elles sont
consid\'erablement plus grandes que les cellules que l'on a utilis\'ees
jusqu'\`a maintenant et qui avaient la taille minimale concevable. On peut
donc les appeller ``cellules \`a gros grains'' ou grosses cellules,
r\'eservant le nom de cellules de taille fine aux pr\'ec\'edentes.

Il est en effet possible de d\'ecouper l'espace des phases en cellules
$E_1,E_2,\ldots=\{E_\kappa\}_{\kappa=1,\ldots}$ qui forment un {\it=
 pavage} ou
une {\it partition} $\PP$ et qui ont la propri\'et\'e de {\it covariance}.

Leur bords sont constitu\'es par une r\'eunion d'axes des syst\`emes
locaux de coordonn\'ees dont on a parl\'e plus haut~: donc les bords
consistent en des surfaces qui soit se contractent sous l'action de
la dynamique, soit se dilatent. On dira que les fronti\`eres des
cellules de la partition $\PP=E_1,E_2,\ldots$ consistent en une partie
qui se contracte ou ``stable'' et en une partie qui se dilate ou
``instable''. La propri\'et\'e de covariance dit alors que sous l'action
de l'\'evolution les cellules se d\'eforment {\it mais les parties stables
de leur bords \'evoluent de fa{\c c}on \`a terminer comme sous--ensembles de
leur r\'eunion}~: la figure suivante illustre cette propri\'et\'e simple.

\*

\eqfig{275pt}{110pt}{
\ins{-8pt}{100pt}{$s$}
\ins{49pt}{90pt}{$\Delta$}
\ins{110pt}{0pt}{$u$}
\ins{260pt}{0pt}{$u$}
\ins{142pt}{100pt}{$s$}
\ins{258pt}{19pt}{$S\Delta$}}
{555}{Fig.1}

Si on a une telle partition (qui s'appelle {\it partition markovienne})
$\PP$ on peut la raffiner en d'autres qui ont la m\^eme propri\'et\'e de
covariance~: simplement en donnant un entier $T$ et consid\'erant la
partition constitu\'ee par les ensembles $S^{-T} E_{\kappa_{-T}}\cap\ldots
S^T E_{\kappa_T}$ qui, \`a cause de la contraction et de l'expansion de
l'espace lors de l'\'evolution, forment une partition $\PP_T$
dont les cellules deviennent aussi petites que l'on veut en
prenant $T$ assez grand.

Si $F$ est une observable on peut en calculer la valeur moyenne
simplement en consid\'erant une partition markovienne $\PP$ (arbitraire,
car il n'y a pas d'unicit\'e) en construisant la partition $\PP_T$ avec
$T$ assez grand pour que $F$ soit constant dans chaque cellule $C$ de
$\PP_T$ et puis en posant~:

$$\media{F}=\fra{\sum_C P(C) F(C)}{\sum_C P(C)}\Eqa(4)$$
o\`u $P(C)$ est un ``poids'' convenable. Il est construit en choisissant un
point $c\in C$ et en consid\'erant son \'evolution entre $-\tau$ et $\tau$
o\`u $\tau$ est grand mais petit par rapport \`a $T$ (par exemple
$\tau=\fra12T$).

On consid\`ere le point $S^{-\tau}c$ qui est transform\'e en $S^\tau c$ en
un temps $2\tau$. On voit que l'axe, par $S^{-\tau}c$, des coordonn\'ees
qui se dilatent sous l'action de l'\'evolution est dilat\'e, au cours
d'un temps $2\tau$, par un facteur qu'on appelle $\Lambda_{2\tau,i}(c)$~: {\it
alors le poids $P(C)$ peut \^etre choisi \'egal \`a
$\Lambda_{2\tau,i}(c)^{-1}$}.

L'\'equation \equ(4) est la formule qui remplace la distribution
microcanonique hors \'equilibre~: on peut prouver que l'on s'y ram\`ene
sous l'hypoth\`ese chaotique. La question qui se pose est si l'on peut
tirer quelques cons\'equences g\'en\'erales de l'hypoth\`ese chaotique
moyennant l'usage de la repr\'esentation \equ(4)
ci-dessus.\annota{2}{L'expression \equ(4) pour la distribution SRB permet
d'\'eclaircir la r\'epr\'esentation de l'\'evolution comme permutation des
cellules \`a taille fine. On doit imaginer que chaque
\'element $C$ (``cellule \`a gros grain'') de la partition markovienne
$\PP_T$, avec un $T$ tr\`es grand de fa{\c c}on \`a ce que toute
observable $F$ (pertinente pour le comportment macroscopique) reste
constante sur chaque $C$~: pour une repr\'esentation fid\`ele du
mouvement, on imagine que chaque $C$ est quadrill\'e par des cellules
tr\`es petites ``de taille fine'' {\it en nombre proportionnel \`a
$P(C)$}. Par l'\'evolution les cellules de taille fine se
r\'epartissent entre les \'el\'ements $C'$ de $\PP_T$ qui
intersectent $S C$. On fait \'evoluer de la m\^eme fa{\c c}on les
autres cellules fines des \'el\'ements de $\PP_T$~: la th\'eorie des
distributions SRB montre que le nombre des cellules de taille fine qui
viennent se trouver dans chaque $C\in\PP_T$ ne change pas, \`a une
tr\`es bonne aproximation pr\`es; c'est la stationnarit\'e de la
distribution SRB, \;([Ga1]). Alors on peut d\'efinir l'\'evolution des
cellules de taille fine simplement en disant qu'une cellule fine $\delta$
dans $C$ \'evolue dans une des cellules fines qui sont dans la $C'$
qui contient $S\delta$; il faut seulement faire attention \`a ne pas
associer une m\^eme cellule fine de $C'$ \`a deux cellules fines
appartenant \`a diffe\'rentes $C$ (parmi celles telles que $SC\cap
C'\ne\emptyset$)~: on peut s'arranger de fa{\c c}on telle que la
permutation des cellules fines ainsi d\'efinie soit \`a un seul cycle,
car les d\'etails du mouvement \`a l'int\'erieur des cellules $C$ n'ont
pas d'importance parce que les observables qui nous int\'eressent sont
constantes dans les $C$.  Mais la m\^eme construction peut \^etre faite
en rempla{\c c}ant le poids $P(C)$ par $P(C)^\alpha$ avec $\alpha\ne1$~: on
obtient ainsi d'autres distributions stationnaires {\it diff\'erentes}
de la SRB, et on peut m\^eme en construire d'autres, \;([S]), \;([R1]). On
peut repr\'esenter de la m\^eme fa{\c c}on aussi ces autres distributions~:
mais on doit imaginer que les cellules de taille fine que l'on utilise
pour en repr\'esenter une soient {\it diff\'erentes} de celles
utilis\'ees pour repr\'esenter les autres. {\it En fin de compte toutes
les cellules fines ainsi introduites repr\'esentent l'attracteur}. Si
on divise l'espace entier en (beaucoup de) cellules fines, de
fa{\c c}on \`a ce que {\it toutes distributions stationnaires} puissent
\^etre repr\'esent\'ees par une permutation des cellules fines qui se
trouvent dans les $C\in \PP_T$, alors on obtient une
r\'epr\'esentation disc\`ete tr\`es fid\`ele du mouvement. Mais
toutes les cellules ne feront pas partie d'un cycle, car la dynamique
est en g\'en\'eral dissipative et une grande partie d'entre elles ne
reviennent pas sur elles m\^emes mais ``tombent sur l'attracteur'' o\`u,
d\`es lors, elles \'evoluent dans un cycle. La th\'eorie de la distribution
SRB montre que si on consid\`ere un ensemble ouvert dans l'espace des
phases le comportement asymptotique du mouvement de tout point, sauf un
ensemble de volume nul, est bien r\'epr\'esent\'e par la distribution
SRB, ce qui lui fait jouer un r\^ole particulier, au contraire des
autres distributions que l'on peut definir~: c'est-\`a-dire que la grande
majorit\'e (en volume) des cellules fines tombant sur l'attracteur
vont se trouver parmi celles que l'on a associ\'ees aux cycles de la
distribution SRB.}

Dans ce contexte, mentionnons que r\'ecemment on
a r\'eussi \`a d\'eduire une con\-s\'e\-quence qui apparemment \`a un
certain int\'er\^et. On va la formuler pour un syst\`eme
d\'ecrit par une \'equation diff\'erentielle (et donc en temps
continu) $\dot x= f(x)$ qui engendre un flot $t\to S^t x$ dans
l'espace des phases. On suppose aussi {\it que l'\'evolution est
r\'eversible}~: c'est-\`a-dire qu'il y a une transformation isom\'etrique
$I$ de l'espace des phases qui anti-commute avec l'\'evolution~:
$IS^t=S^{-t}I$.

Imaginons un syst\`eme pour lequel l'hypoth\`ese chaotique soit
valable, donc d\'ecrit par une \'equation $\dot x= f(x)$ et soit
$\sigma(x)=-{\rm div}\, f(x)$ la contraction de l'espace des phases
associ\'ee. Supposons que l'on mesure la quantit\'e $\sigma(S^nx)$ au cours
du temps mais avec le syst\`eme dans son \'etat stationnaire. Appellons
$\sigma_+$ sa moyenne temporelle {\it que l'on suppose non nulle} (alors
elle ne peut \^etre que positive par un th\'eor\`eme g\'en\'eral,
\;([R3])) et~:

$$p=\fra1{\tau \sigma_+}\ig_{-\fra12\tau}^{\fra12\tau} \sigma(S^t x)dt$$
et soit $\pi_\tau(p)=e^{\tau\zeta(p)}$ la distribution de
probabilit\'e de cette observable. Alors~:

$$\fra{\zeta(p)-\zeta(-p)}{\tau\sigma_+}\=1$$
c'est le {\it th\'eor\`eme de fluctuation}, \;([GC]).

Je ne peux pas discuter ici la signification physique du th\'eor\`eme
et de l'hypoth\`ese de r\'eversibilt\'e, mais il est int\'eressant de
souligner sa g\'en\'eralit\'e, son ind\'ependance du syst\`eme consid\'er\'e
et aussi l'absence, dans sa formulation, de param\`etres libres. Ce
qui le rend en un certain sens analogue au th\'eor\`eme de la chaleur,
qui lui aussi est g\'en\'eral et sans param\`etres libres.

Il suffira de dire que le th\'eor\`eme de fluctuation est une
propri\'et\'e qu'il faut quand m\^eme v\'erifier exp\'erimentalement~: en
effet une partie de la th\'eorie ci-dessus est n\'ee \`a la suite d'une
exp\'erience de simulation num\'erique et pour en interpr\'eter
th\'eoriquement les r\'esultats, [ECM]. Il y a eu aussi quelques
v\'erifications ind\'ependantes, [BGG], [LLP].

La raison pour laquelle des exp\'eriences sont n\'ecessaires est qu'il
n'y a aucun espoir de prouver que des syst\`emes r\'eels v\'erifient au
sens math\'ematique du mot l'hypoth\`ese chaotique; moins encore
de prouver que des syst\`emes r\'eel v\'erifient l'hypoth\`ese
ergodique. Il n'y a m\^eme pas d'espoir de prouver que des
syst\`emes int\'eressants en simulation num\'erique ou dans la r\'ealit=E9
v\'erifient des propri\'et\'es qui soient assez proches de celles des
syst\`emes hyperboliques pour en d\'eduire des cons\'equences telles que le
th\'eor\`eme de fluctuation. Mais on peut croire que n\'eanmoins ``les
choses se passent comme si l'hypoth\`ese chaotique \'etait litt\'eralement
vraie''.

Il y a donc une n\'ecessit\'e d'un contr\^ole exp\'erimental car on est dans
la m\^eme situation qu'\`a l'\'equilibre~: o\`u tout en croyant, avec
=46eynman, que ``{\it if we follow our solution} [\ie motion] {\it for a lon=
g
enough time it tries everything that it can do, so to speak}'' (see
p. 46-4/5 in [F], vol. I), il a \'et\'e n\'eanmoins n\'ecessaire de faire
de bonnes v\'erifications exp\'erimentales pour ne plus avoir de r\'eserves
ou de doutes sur l'hypoth\`ese ergodique dans la th\'eorie de l'\'equilibre.

Quelques r\'ef\'erences sont donn\'ees ici pour guider le lecteur dans la
litt\'erature r\'ecente et ancienne. Elles sont loin d'\^etre exhaustives~:
[WW], [ZZ].

\bigskip
{
\0\it%
Bibliographie.\rm
\*
\advance\leftskip 1cm

\item{[B66]} Boltzmann, L.: {\it\"Uber die mechanische Bedeutung des
zweiten Haupsatzes der W\"arme\-theo\-rie}, in
``Wissenschaftliche Abhandlungen'', ed. F. Hasen\"ohrl,
vol. I, p. 9--33, reprinted by Chelsea, New York.

\item{[B68]} Boltzmann, L.: {\it Studien \"uber das Gleichgewicht der
lebendigen Kraft zwischen bewegten materiellen Punkten}, in
``Wissenschaftliche Abhandlungen'', ed. F. Hasen\"ohrl,
vol. I, p.  49--96, reprinted by Chelsea, New York.

\item{[B71]} Boltzmann, L.: {\it Analytischer Beweis des zweiten Hauptsatzes
der mechanischen W\"arme\-theo\-rie aus den S\"atzen \"uber das
Gleichgewicht des lebendigen Kraft}, in ``Wissenschaftliche
Abhandlungen'', ed. F. Hasen\"ohrl, vol. I, p.  288--308, reprint\-ed by
Chelsea, New York.

\item{[B72]} Boltzmann, L.: {\it Weitere Studien \"uber das
W\"armegleichgewicht unter Gasmolek\"u\-len}, english translation in S.
Brush, {\it Kinetic theory}, {Vol. 2}, p. 88. Original in
``Wissenschaftliche Abhandlungen'', ed. F. Hasen\"ohrl,
vol. I, p. 316--402, reprinted by Chelsea, New York.

\item{[B77]} Boltzmann, L.: {\it \"Uber die Beziehung zwischen dem
zweiten Hauptsatze der mechanischen W\"armetheorie und der
Wahrscheinlichkeitsrechnung, respektive den S\"atz\-en \"uber das
W\"ar\-me\-gleich\-ge\-wicht}, in "Wis\-sen\-schaft\-li\-che
Abhandlungen'', vol. II, p. 164--223, F. Hasen\"ohrl, Chelsea, New
York, 1968 (reprint).

\item{[B84]} Boltzmann, L.: {\it \"Uber die Eigenshaften monozyklischer
und anderer damit verwandter Systeme}, in ``Wissenschaftliche
Abhandlungen'', ed. F.P. Hasen\"ohrl, vol. III,
Chelsea, New York, 1968, (reprint).

\item{[B96]} Boltzmann, L.: {\it Entgegnung auf die
w\"armetheoretischen Betrachtungen des Hrn.  E.  Zermelo}, engl.
transl.: S.  Brush, "Kinetic Theory", {vol. 2}, 218, Pergamon
Press.

\item{[B97]} Boltzmann, L.: {\it Zu Hrn.  Zermelo's Abhandlung "Ueber
die mechanische Er\-kl\"a\-rung irreversibler Vorg\"ange}, engl.
trans. in S.  Brush, "Kinetic Theory", {vol. 2}, 238, Pergamon
Press.

\item{[B02]} Boltzmann, L.: {\it Lectures on gas theory}, english
edition annotated by S.  Brush, University of California Press,
Berkeley, 1964.

\item{[B05]} Boltzmann, L.: {\it More on atomism} inclus dans {\it
Theoretical physics and philosophical problems}, p.54--56,
ed. B. McGuinness, Reidel, 1974, traduction de {\it Popul\"are
Schriften}, ed. J.A. Barth, Leipzig, 1905.

\item{[BR]} R.Bowen, D.Ruelle. {\it Ergodic theory of Axiom A flows},
Inventiones Math. {\bf 2},181-202(1975).  See also Ruelle, D.: {\it Chaotic
motions and strange attractors}, Lezioni Lincee, notes by S.  Isola,
Accademia Nazionale dei Lincei, Cambridge University Press, 1989.

\item{[BGG]} Bonetto, F., Gallavotti, G.,
Garrido, P.: {\it Chaotic principle: an experimental test}, Physica D,
{\bf 105}, 226--252, 1997,

\item{[ECM]} Evans, D.J.,Cohen, E.G.D., Morriss, G.P.: {\it
Probability of second law violations in shearing steady flows},
Phys. Rev.  Letters, {\bf 71}, 2401--2404, 1993.

\item{[G]} Gibbs, J.: {\it Elementary principles in statistical
mechanics}, Ox Bow Press, 1981, (reprint).

\item{[Ga1]} Gallavotti, G.: {\sl Meccanica Statistica},
``Quaderni del CNR-GNFM'', vol. {\bf 50}, p. 1--350, Firenze, 1995.

\item{[Ga2]} Gallavotti, G.: {\it Ergodicity, ensembles, irreversibility
in Boltzmann and beyond}, Journal of Statistical Physics, {\bf 78},
1571--1589, 1995.

\item{[GC]} Gallavotti, G., Cohen, E.G.D.: {\it Dynamical
ensembles in non-equilibrium statistical mechanics}, Physical Review
Letters, {\bf74}, 2694--2697, 1995. Gallavotti, G., Cohen,
E.G.D.: {\it Dynamical ensembles in stationary states}, Journal of
Statistical Physics, {\bf 80}, 931--970, 1995.

\item{[H]} Helmholtz, H.: {\it Principien der Statik monocyklischer
Systeme}, in ``Wissenschaft\-li\-che Abhandlungen'', vol.  III, p.  142--162
and p.  179-- 202, Leipzig, 1895. And {\it Studien zur Statik monocyklischer
Systeme}, in ``Wissenschaftliche Abhandlungen'', vol.  III, p.  163--172
and p.  173-- 178, Leipzig, 1895.

\item{[LLP]} Lepri. S., Livi, R., Politi, A.  {\it Energy transport in
anharmonic lattices close and far from equilibrium}, preprint,
cond-mat@xyz. lanl. gov \#9709156.

\item{[R1]} Ruelle, D.: {\it Measures describing a turbulent flow},
Annals of the New York Academy of Sciences, {\bf 357}, 1--9, 1980.

\item{[R2]} Ruelle, D.: {\it A measure associated with Axiom A
attractors}, American Journal of Mathematics, {\bf98}, 619--654, 1976.

\item{[R3]} Ruelle, D.: {\it Positivity of entropy
production in nonequilibrium statistical mechanics}, Journal of
Statistical Physics, {\bf 85}, 1--25, 1996. Et {\it Positivity of
entropy production in the presence of a random thermostat}, Journal of
Statistical Physics, {\bf 86}, 935--951, 1997.

\item{[S]} Sinai, Y.G.: {\sl Gibbs measures in ergodic theory},
Russian Math. Surveys, {\bf 27}, 21--69, 1972.  Also: {\it
Introduction to ergodic theory}, Prin\-ce\-ton U.  Press, 1977.

\smallbreak
\item{[WW]} Il est impossible ici de discuter analytiquement
la vaste litt\'erature qui suivit [R2] et pr\'ec\'eda [GC]. Les travaux
suivants sont remarquables et contiennent des r\'ef\'erences
d\'etaill\'ees \`a d'autres travaux, aussi remarquables:
\\
Holian, B.L., Hoover, W.G., Posch. H.A.:
{\it Resolution of Loschmidt's paradox: the origin of irreversible
behavior in reversible atomistic dynamics}, Physical Review Letters,
{\bf 59}, 10--13, 1987.
\\
Evans, D.J., Morriss, G.P.: {\it Statistical
Mechanics of Non-equilibrium fluids}, Academic Press, New York, 1990.
\\
Evans, D.J.,Cohen, E.G.D., Morriss, G.P.: {\it Viscosity of a simple
fluid from its maximal Lyapunov exponents}, Physical Review, {\bf
42A}, 5990--\-5997, 1990.
\\
Dellago, C., Posch, H., Hoover, W.: {\it Lyapunov instability in
system of hard disks in equilibrium and non-equilibrium steady
states}, Physical Review, {\bf 53E}, 1485--1501, 1996.

\smallbreak\item{[ZZ]} La litt\'erature qui a suivi \`a [GC] est aussi
vaste mais une liste assez compl\`ete est la suivante:
\\
Gallavotti, G.: {\it Reversible Anosov maps and large deviations},
Mathematical Physics Electronic Journal, MPEJ, (http://
mpej.unige.ch), {\bf 1}, 1--12, 1995.
\\
Gallavotti, G.: {\it Chaotic hypothesis: Onsager reciprocity and
fluctuation--dissipation theorem}, Journal of Statistical Phys., {\bf
84}, 899--926, 1996.
\\
Gallavotti, G.: {\it Extension of Onsager's's reciprocity
to large fields and the chaotic hypothesis}, Physical Review Letters,
{\bf 77}, 4334--4337, 1996.
\\
Gallavotti, G.: {\it Chaotic principle: some applications to developed
turbulence}, Journal of Statistical Physics, {\bf 86}, 907--934, 1997.
\\
Gentile, G: {\it Large deviation rule for Anosov flows}, {\tt mp$\_$arc@
math.utexas.edu, \#96--79}, in print in Forum Mathematicum.
\\
Ruelle, D.: {\it Differentiation of SRB states},
Communications in Mathematical Physics, {\bf 187}, 227--241, 1997.
\\
Gallavotti, G., Ruelle, D.: {\it SRB states and non-equilibrium
statistical mechanics close to equilibrium}, Communications in
Mathematical Physics, in print; archived in {\tt mp$\_$arc@
math.utexas.edu, \# 96--645}.
\\
Gallavotti, G.: {\it Fluctuation patterns and conditional
reversibility in non-equilibrium systems}, in print on Annales de
l'Institut H. Poincar\'e, {\tt chao-dyn@xyz.lanl.gov \#9703007}.
\\
Gallavotti, G.: {\it New methods in non-equilibrium gases
and fluids}, Proceedings of the conference {\sl Let's face chaos
through nonlinear dynamics}, U. of Maribor, 24 June-- 5 July 1996,
ed. M. Robnik.
\\
Gallavotti, G.: {\it Equivalence of dynamical
ensembles and Navier Stokes equations}, Physics Letters, {\bf223A},
91--95, 1996.
\\
Gallavotti, G.: {\it Dynamical ensembles equivalence in
fluid mechanics}, Physica D, {\bf 105}, 163--184, 1997.
\\
Ruelle, D.: {\it Entropy production in nonequilibrium statistical
mechanics}, Communications in Mathematical Physics,
{\bf189}, 365--371, 1997.
\\
Ruelle, D.: {\it New theoretical ideas in non-equilibrium statistical
mechanics}, Lecture notes at Rutgers University, fall 1997.
\\
Bonetto, F., Gallavotti, G.: {\it Reversibility, coarse graining and
the chaoticity principle} Communications in Mathematical Physics,
{\bf189}, 263--276, 1997.
\\
Morriss, G.P., Rondoni, L.:
{\it Applications of periodic orbit theory to $N$--particle systems},
Journal of Statistical Physics, {\bf 86}, 991, 1997.
\\
Kurchan, J.: {\it Fluctuation Theorem for stochastic dynamics},
preprint, cond-mat@xyz. lanl. gov, \# 9709304.

\*
\0{\it Internet:
Les derni\`eres versions des  pr\'etirages de l'Auteur se trouvent aussi
\`a~:

\centerline{\tt http://chimera.roma1.infn.it/}
\centerline{\tt http://www.math.rutgers.edu/$\sim$giovanni/}
\par
\0Mathematical Physics Preprints (mirror) pages.\par

\medskip
\sl e-mail: \tt gallavotti@roma1.infn.it
}}

\ciao